\documentclass[copyright]{eptcs}
\providecommand{\event}{TARK 2017} 
\usepackage{underscore}           
\usepackage[pdftex]{graphicx}
\usepackage{amssymb}
\usepackage{amsmath}
\usepackage{amsfonts}
\usepackage[english]{babel}
\usepackage{bm}

\newtheorem{defin}{Definition}

\newtheorem{prop}{Proposition}

\newtheorem{theo}{Theorem}

\title{Self-Confirming Games: Unawareness, Discovery, and Equilibrium}

\author{Burkhard C. Schipper
\institute{Department of Economics \\
University of California, Davis}
\email{bcschipper@ucdavis.edu}}
\def\titlerunning{Self-Confirming Games: Unawareness, Discovery, and Equilibrium}
\def\authorrunning{Burkhard C. Schipper}

\begin{document}

\maketitle

\begin{abstract} Equilibrium notions for games with unawareness in the literature cannot be interpreted as steady-states of a learning process because players may discover novel actions during play. In this sense, many games with unawareness are ``self-destroying'' as a player's representation of the game must change after playing it once. We define discovery processes where at each state there is an extensive-form game with unawareness that together with the players' play determines the transition to possibly another extensive-form games with unawareness in which players are now aware of actions that they have previously discovered. A discovery process is rationalizable if players play extensive-form rationalizable strategies in each game with unawareness. We show that for any game with unawareness there is a rationalizable discovery process that leads to a self-confirming game that possesses an extensive-form rationalizable self-confirming equilibrium. This notion of equilibrium can be interpreted as steady-state of a learning and discovery process.
\end{abstract}

\section{Introduction}

How do players arrive at their conception(s) of a strategic situation? Are representations of strategic situations necessarily common among all players? How to model discoveries of novel actions? What is ``equilibrium'' in games with unawareness? These are the questions I attack in this paper. In particular, our motivation is the quest for a natural notion of equilibrium to games with unawareness. Various frameworks for modeling dynamic games with unawareness have been recently introduced (Halpern and Rego, 2014, Rego and Halpern, 2012, Feinberg, 2012, Li, 2008, Grant and Quiggin, 2013, Heifetz, Meier, and Schipper, 2013). While all of those frameworks are capable of modeling strategic interaction under asymmetric unawareness at various degrees of generality and tractability, the solution concepts proposed for those frameworks and thus the implicit behavioral assumptions under unawareness differ. They can roughly be divided into equilibrium notions (Halpern and Rego, 2014, Rego and Halpern, 2012, Feinberg, 2012, Li, 2008, Grant and Quiggin, 2013, Meier and Schipper, 2013) and rationalizability notions (Heifetz, Meier, and Schipper, 2013, 2012, Meier and Schipper, 2012). Authors proposing equilibrium notions to dynamic games with unawareness appear to be mainly guided by extending the mathematical definitions of equilibrium in standard games to the more sophisticated frameworks with unawareness. Yet, I believe less attention has been paid to the interpretations of the behavioral assumptions embodied in these standard equilibrium concepts and whether or not such interpretations apply also to dynamic games with unawareness.

In standard game theory, equilibrium is interpreted as an outcome in which each player plays ``optimally'' given the opponents' play that could have emerged in a steady-state of some learning process. This interpretation cannot apply generally to game with unawareness. This is because players may be unaware of actions and may discover novel actions during play. The ''next time'' they play ``the game'', they actually play a different game in which now they are aware of previously discovered actions. That is, dynamic learning processes in games with unawareness must not only deal with learning about opponents' play but also with discoveries that may lead to changes in players' representations of the game. Games with unawareness may be ``self-destroying'' representations of the strategic situation in the sense that (rational) play may destroy some player's representation of the strategic situation. Only when a representation of the strategic situation is ``self-confirming'', i.e., (rational) play in such a game does not lead to (further) changes in the players' representation of the game, an equilibrium notion as a steady-state of a learning process may be applied. Our paper makes this precise.

We introduce a notion of self-confirming equilibrium to extensive-form games with unawareness. In self-confirming equilibrium players play in a way that nobody discovers that their own view of the game may be incomplete. Moreover, players play optimally given their beliefs and their beliefs are not falsified given the play. We show that such a self-confirming equilibrium may fail to exist in an extensive-form game with unawareness because rational play may lead to discoveries. We formalize the notion of discovered game: For any extensive-form game with unawareness and strategy profile, the discovered game is a game in which each player's awareness is ``updated'' given their discoveries but their information stays essentially the same (modulo awareness). This leads to a notion of a stochastic game in which states correspond to extensive-form games with unawareness and the transition probabilities model for each extensive-form game with unawareness and strategy profile the transition to the discovered game. Such a stochastic game together with a Markov strategy for each player that assigns to each extensive-form game with unawareness a mode of behavior we call a discovery process. We select among discovery processes by requiring the stochastic games Markov strategy to assign only rationalizable strategies to each extensive-form game with unawareness. We show that for every finite extensive-form game with unawareness, there exists an extensive-form rationalizable discovery process that leads to a extensive-form game with unawareness that is an absorbing state of the process. We consider it as a steady state of conceptions when players play with common (strong) belief in rationality and call it rationalizable self-confirming game. In such a game, it makes sense to look also for a steady state of behavior by focusing on self-confirming equilibrium involving only extensive-form rationalizable strategies. We show that for every extensive-form game with unawareness there exists a rationalizable discovery process leading to a rationalizable self-confirming game that possesses a rationalizable self-confirming equilibrium. This is a notion of equilibrium both in terms of conceptions of the strategic situation as well as strategic behavior.

Besides to the aforementioned literature on games with unawareness, the paper is related to the literature on self-confirming equilibrium in games (Battigalli, 1987, Fudenberg and Levine, 1993, Kalai and Lehrer, 1993, Fudenberg and Kreps, 1995, Battigalli et al., 2015) in particular to papers using some related notions of rationalizable self-confirming equilibrium (Rubinstein and Wolinski, 1994, Dekel, Fudenberg, and Levine, 1999, 2002, Fudenberg and Kamada, 2015, 2016, Gilli, 1999, Esponda, 2013). Finally, Greenberg, Gutpa, and Luo (2009), Sasaki (2016), and Copic and Galeotti (2006) combine ideas of self-confirming equilibrium and lack of awareness although none present explicit models of discoveries.

Our notion of discovered game can be understood as a game theoretic analogue to awareness bisimulation in van Ditmarsch et al. (2013). Awareness bisimulation is used to compare awareness structures of Fagin and Halpern (1988). Roughly, two awareness structures are awareness bisimilar if they model the same information modulo awareness. In our context, the discovered version of a game with unawareness models the same information modulo awareness.

\section{Simple Example\label{Example}}

\begin{figure}[h!]
\caption{(Initial) Game of Example 1}
\label{example1a}
\begin{center}
\includegraphics[scale=0.4]{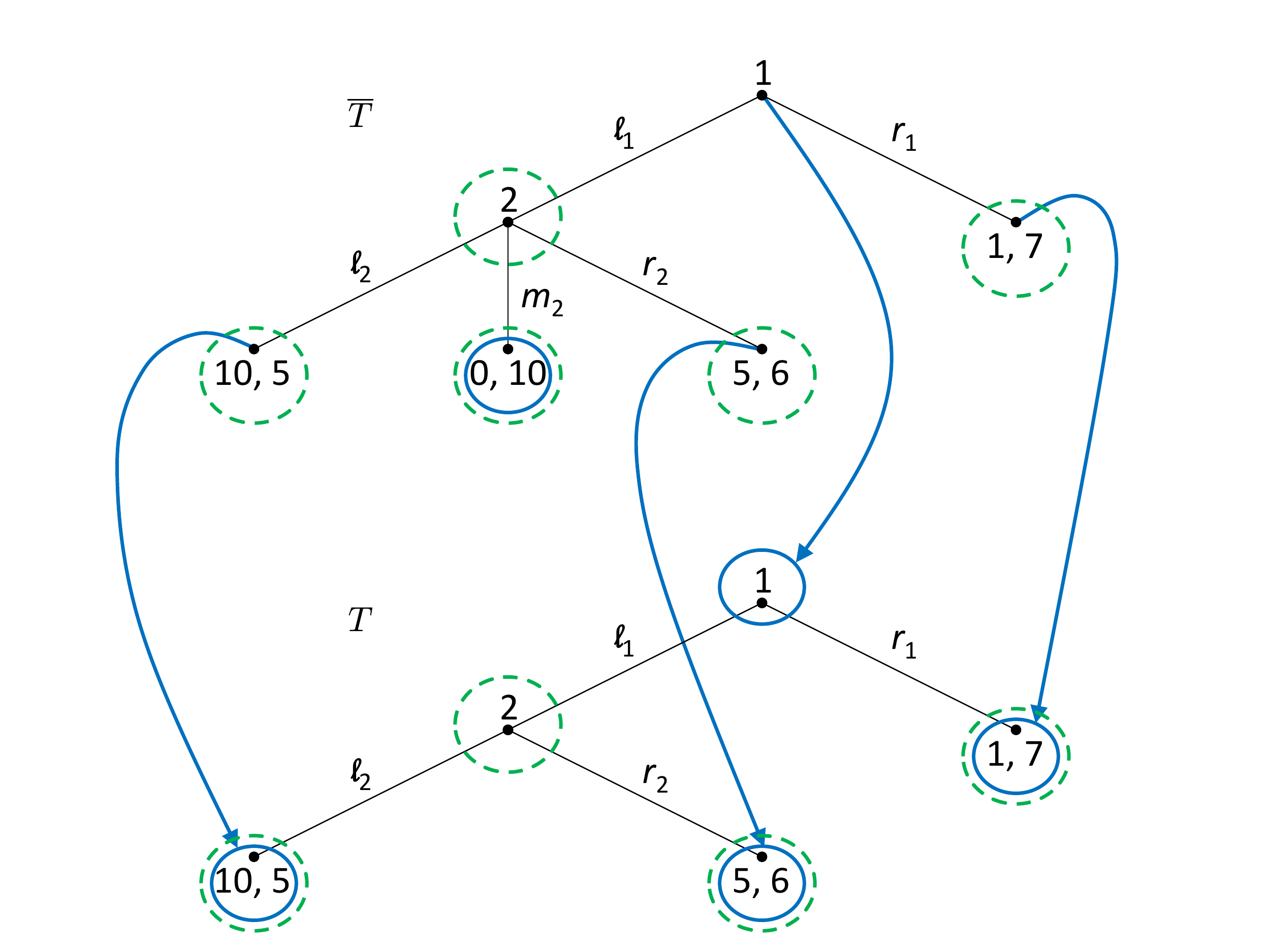}
\end{center}
\end{figure}

There are two players, 1 and 2. Player 1 moves first. She can either delegate to player 2 or do the work by herself. In the latter case, the game ends and both players receive their payoffs. If player 1 delegates to player 2, then player 2 can take one out of three actions. The non-standard but straightforward detail is that player 1 is not aware of all of player 2's actions. She considers only two actions of player 2. This strategic situation is modeled in the game depicted in Figure~\ref{example1a}.

There are two trees. The tree at the bottom, $T$, is a subtree of the tree at the top, $\bar{T}$, in the sense that action $m_2$ of player 2 is missing in $T$. The information and awareness of both players is modeled with information sets. The solid-lined blue spheres and arrows belong to player 1, the dashed green spheres belong to player 2. There are two non-standard features of these information sets. First, the information set of a decision node in $\bar{T}$ may consist of decision nodes in a lower tree $T$. For instance, player 1's information set at the beginning of the game in the upper tree $\bar{T}$ is in the lower tree $T$. This signifies the fact that initially player 1 is unaware of player 2's action $m_2$ and thus considers the strategic situation to be represented by the tree at the bottom, $T$. Second, we added information sets at terminal nodes. The reason is that in order to discuss notions of equilibrium under unawareness, it will be useful to analyze also the players' views at the end of the game. As usual, the information in extensive-form games is represented by information sets. Players receive a payoff at each terminal node. The first component at each terminal node refers to player 1's payoff whereas the second component refers to player 2's payoff.

\begin{figure}[h!]
\caption{Game of Example 1 after being played once}
\label{example1b}
\begin{center}
\includegraphics[scale=0.4]{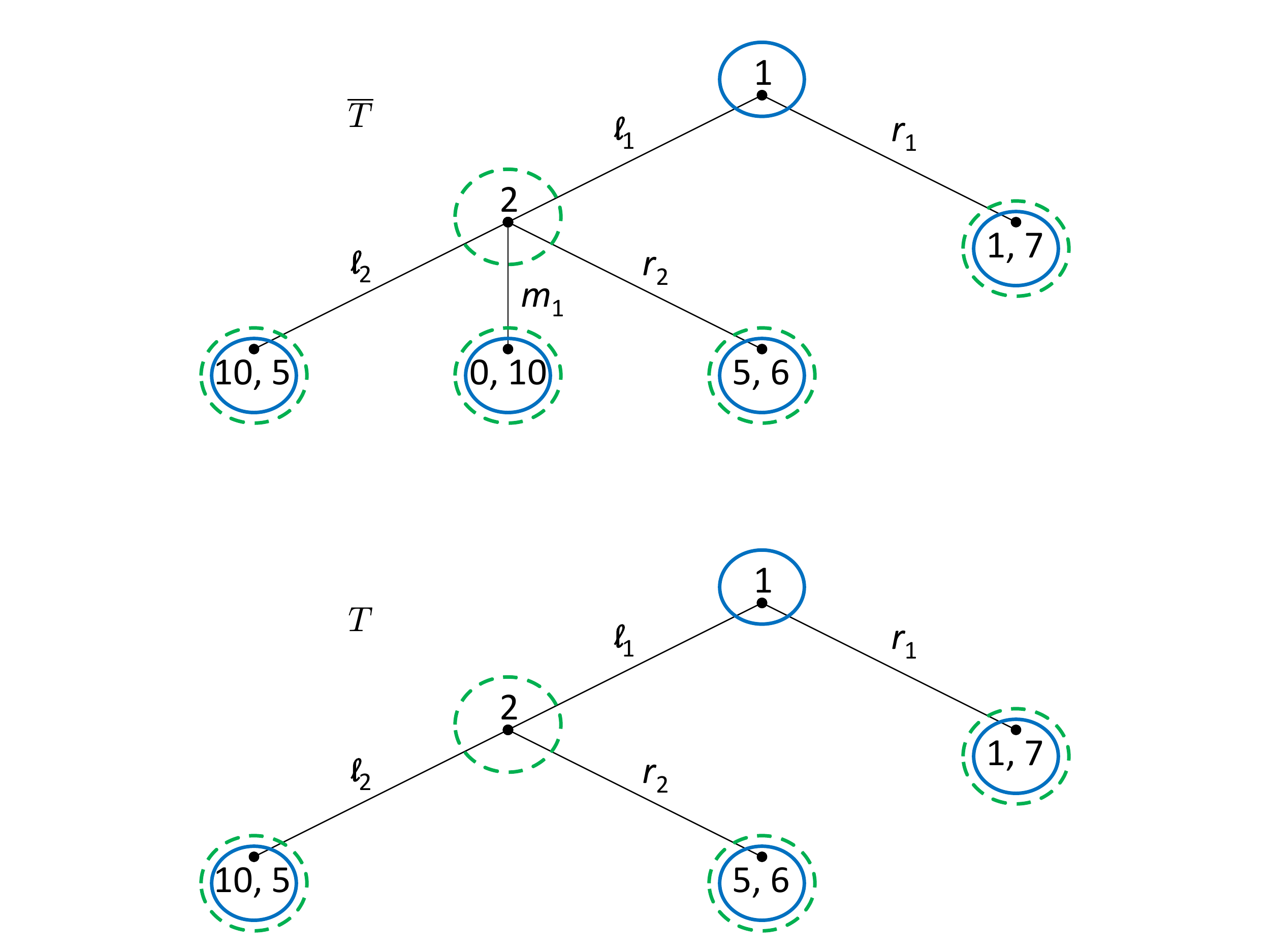}
\end{center}
\end{figure}

What is equilibrium in this game? A basic requirement is that in equilibrium players should play rational. That is, each player at each information set where (s)he is to move should play an action that maximizes her expected payoff subject to her belief over the opponent's behavior. At the beginning of the game, player 1 thinks that she faces the situation depicted in tree $T$. Clearly, with this mindset only action $\ell_1$ is rational because no matter what she expects player 2 to do, she obtains a higher expected payoff from playing $\ell_1$ than from $r_1$. At the information set in the upper tree $\bar{T}$, player 2 is aware of his action $m_2$. Since $m_2$ strictly dominates any of his other actions, the only rational action for player 2 at this information set is to choose $m_2$. Thus, the path of play emerging from any rational play is $(\ell_1, m_2)$ with player 1 obtaining zero payoff and player 2 obtaining a payoff of $10$. Yet, we strongly believe that no profile of rational strategies can be reasonably called an equilibrium in this setting because any profile of strategies in which player 1 chooses $\ell_1$ and player 2 chooses $m_2$ cannot be interpreted as a steady state of a learning process. After players choose rationally in the game, player 1's awareness changed. She discovered action $m_2$ of player 2. This is symbolized by player 1's information set at the terminal node after $m_2$ in the tree $\bar{T}$. Thus, the ``next'' time players do not play the game of Figure~\ref{example1a} but a ``discovered version'' of it in which player 1 is aware of action $m_2$ upfront. This discovered game is depicted in Figure~\ref{example1b}. At the beginning of the game, player 1's information set is now in the upper tree $\bar{T}$. Consequently she is aware of all actions of all players. She won't be surprised by any terminal node as her information sets at terminal nodes in the upper tree $\bar{T}$ also lie in this tree. The lower tree $T$ becomes in some sense redundant as players are now commonly aware of the strategic situation modeled by $\bar{T}$. Yet, since they are aware, they can envision themselves also in a situation in which both players are unaware of $m_2$, which is what now $T$ represents although this counterfactual mindset is not behaviorally relevant.

In the discovered version shown in Figure~\ref{example1b}, the only rational action for player 1 at the beginning of the game is to choose $r_1$ in $\bar{T}$. Thus any steady state of a learning (and discovery) process must prescribe $r_1$ for player 1 in $\bar{T}$.

To sum up, we note first that games with unawareness may not possess equilibria that can be interpreted as steady states of a learning process (see the game in Figure~\ref{example1a}). Second, an equilibrium notion capturing the idea of a steady-state of a learning (and discovery) process in games with unawareness must not only involve usual conditions on behavior of players but must also impose restrictions on representations of the strategic situation. That is, their representations of the strategic situation must be consistent with their behavior and behavior must be consistent with their representations of the strategic situations. To emphasize this, we will use the terminology of self-confirming games.

\section{Games with Unawareness\label{model}}

We outline extensive-form game with unawareness as introduced by Heifetz, Meier, and Schipper (2013) together with some crucial extensions required for our analysis. To define a extensive-form game with unawareness $\Gamma$, consider first, as a building block, a finite game with perfect information and possibly simultaneous moves. This tree is there to outline all physical moves. There is a finite set of players $I$ and possibly a special player ``nature'' with index $0$. We denote by $I^0$ the set of players including nature. Further, there is a nonempty finite set of ``decision'' nodes $\bar{D}$ and a player correspondence $P: \bar{D} \longrightarrow 2^{I^0} \setminus \{\emptyset\}$ that assigns to each node $n \in \bar{D}$, a nonempty set of ``active'' players $P(n) \subseteq I^0$. (That is, we allow for simultaneous moves.) For every decision node $n \in \bar{D}$ and player $i \in P(n)$ who moves at that decision node, there is a nonempty finite set of actions $A_{n}^{i}$. Moreover, there is a set of terminal nodes $\bar{Z}$. Since we will also associate information sets with terminal nodes for each player, it will be useful to extent $P$ to $\bar{Z}$ by $P(z) = I$ and let $A_z^i = \emptyset$ for all $i \in I$, $z \in \bar{Z}$. Finally, each terminal node $z \in \bar{Z}$ is associated with a vector of payoffs $(u_i(z))_{i \in I}$. We require that nodes in $\bar{N} := \bar{D} \cup \bar{Z}$ constitute a tree denoted by $\bar{T}$. That is, nodes in $\bar{N}$ are partially ordered by a precedence relation $\lessdot$ with which $(\bar{N}, \lessdot)$ forms an arborescence (that is, the predecessors of each node in $\bar{N}$ are totally ordered by $\lessdot$), there is a unique node in $\bar{N}$ with no predecessors (i.e., the root of the tree), for each decision node $n \in \bar{D}$ there is a bijection $\psi_n$ between the action profiles $\prod_{i \in P(n)} A_{n}^{i}$ at $n$ and $n$'s immediate successors, and any terminal node in $\bar{Z}$ has no successors.

Note that so far we treat nature like any other player except that at terminal nodes we do not assign payoffs to nature. We do not need to require that nature moves first or that nature moves according a pre-specified probability distribution (although these assumptions can be imposed in our framework).

Consider now a join-semilattice $\mathbf{T}$ of subtrees of $\bar{T}$. A subtree is defined by a subset of nodes $N \subseteq \bar{N}$ for which $(N, \lessdot)$ is also a tree (i.e., an arborescence in which a unique node has no predecessors). Two subtrees $T', T'' \in \mathbf{T}$ are ordered, written $T' \preceq T''$ if the nodes of $T'$ constitute a subset of the nodes of $T''$. We require three properties:
\begin{enumerate}
\item All the terminal nodes in each tree $T \in \mathbf{T}$ are in $\bar{Z}$. That is, we don't create ``new'' terminal nodes.

\item For every tree $T \in \mathbf{T}$, every node $n \in T$, and every active player $i \in P(n)$ there exists a nonempty subset of actions $A_{n}^{i,T} \subseteq A_{n}^{i}$ such that $\psi_{n}$ maps the action profiles $A_{n}^{T} = \prod_{i \in P(n)} A_{n}^{i,T}$ bijectively onto $n$'s successors in $T$.

\item If for two decision nodes $n, n' \in T$ with $i \in P(n) \cap P(n')$ it is the case that $A_{n}^{i} \cap A_{n'}^{i} \neq \emptyset $, then $A_{n}^{i} = A_{n'}^{i}$.
\end{enumerate}

Within the family $\mathbf{T}$ of subtrees of $\bar{T}$, some nodes $n$ appear in several trees $T \in \mathbf{T}$. In what follows, we will need to designate explicitly appearances of such nodes $n$ in different trees as distinct entities. To this effect, in each tree $T \in \mathbf{T}$ label by $n_{T}$ the copy in $T$ of the node $n \in \bar{N}$ whenever the copy of $n$ is part of the tree $T$, with the requirement that if the profile of actions $a_{n} \in A_{n}^{T}$ leads from $n$ to $n'$, then $a_{n_T}$ leads also from the copy $n_{T}$ to the copy $n_{T}'$. More generally, for any $T, T', T'' \in \mathbf{T}$ with $T \preceq T' \preceq T''$ such that $n \in T''$, $n_{T'}$ is the copy of $n$ in the tree $T'$, $n_T$ is the copy of $n$ in the tree $T$, and $(n_{T'})_{T}$ is the copy of $n_{T'}$ in the tree $T$, we require that ``nodes commute'', $n_T = (n_{T'})_T$. For any $T \in \mathbf{T}$ and any $n \in T$, we let $n_T := n$.

Denote by $\mathbf{D}$ the union of all decision nodes in all trees $T \in \mathbf{T}$, by $\mathbf{Z}$ the union of terminal nodes in all trees $T \in \mathbf{T}$, and by $\mathbf{N} = \mathbf{D} \cup \mathbf{Z}$. Copies $n_{T}$ of a given node $n$ in different subtrees $T$ are now treated distinct from one another, so that $\mathbf{N}$ is a disjoint union of sets of nodes.

In what follows, when referring to a node in $\mathbf{N}$ we will typically avoid the subscript indicating the tree $T$ for which $n \in T$ when no confusion arises. For a node $n \in \mathbf{N}$ we denote by $T_{n}$ the tree containing $n$.\footnote{Bold capital letters refer to sets of elements across trees.}

Denote by $N^{T}$ the set of nodes in the tree $T \in \mathbf{T}$. Similarly, denote by $D_i^T$ the set of decision nodes in which player $i$ is active in the tree $T \in \mathbf{T}$. Finally, denote by $Z^T$ the set of terminal nodes in the tree $T \in \mathbf{T}$.

Information sets model both information and awareness. At a node $n$ of the tree $T_{n} \in \mathbf{T}$, the player may conceive the feasible paths to be described by a different (i.e., less expressive) tree $T' \in \mathbf{T}$. In such a case, her information set will be a subset of $T'$ rather than of $T_{n}$ and $n$ will not be contained in the player's information set at $n$.

In order to define a notion of self-confirming equilibrium we also need to consider the player's view at terminal nodes. Thus we will also devise information sets of terminal nodes that model both the player's information and awareness at the ends of the game. This is different from Heifetz, Meier, and Schipper (2013) but akin to signal, outcome, or feedback functions in some works on self-confirming equilibrium, see for instance Battigalli and Guaitoli (1997) and Battigalli et al. (2015).

Formally, for each node $n \in \mathbf{N}$ (including terminal nodes in $\mathbf{Z}$), define for each active player $i \in P(n)$ a nonempty information set $h_i(n)$ with the following properties:

\begin{itemize}
\item[U0] Confined awareness: If $n \in T$ and $i \in P(n)$, then $h_i(n)
\subseteq T'$ with $T' \preceq T$.

\item[U1] Generalized reflexivity: If $T' \preceq T$, $n \in T$, $h_i(n) \subseteq T'$ and $T'$ contains a copy $n_{T'}$ of $n$, then $n_{T'} \in h_i(n)$.

\item[I2] Introspection: If $n' \in h_i(n)$, then $h_i(n') = h_i(n)$.

\item[I3] No divining of currently unimaginable paths, no expectation to forget currently conceivable paths: If $n' \in h_i(n) \subseteq T'$ (where $T' \in \mathbf{T}$ is a tree) and there is a path $n', \dots , n'' \in T'$ such that $i \in P(n') \cap P(n'')$, then $h_i(n'') \subseteq T'$.

\item[I4] No imaginary actions: If $n' \in h_i(n)$, then $A_{n'}^i \subseteq A_n^i$.

\item[I5] Distinct action names in disjoint information sets: For a subtree $T \in \mathbf{T}$, if there a decision nodes $n, n' \in T \cap \mathbf{D}$ with $A_n^i = A_{n'}^i$, then $h_i(n') = h_i(n)$.

\item[I6] Perfect recall: Suppose that player $i$ is active in two distinct nodes $n_{1}$ and $n_{k}$, and there is a path $n_{1}, n_{2}, ..., n_{k}$ such that at $n_{1}$ player $i$ takes the action $a_{i}$. If $n^{\prime} \in h _{i}\left( n_{k}\right)$, $n' \neq n_k$, then there exists a node $n_{1}^{\prime}\neq n^{\prime }$ and a path $n_{1}^{\prime}, n_{2}^{\prime }, ..., n_{\ell}^{\prime } = n^{\prime }$ such that $h_{i}\left( n_{1}^{\prime}\right) = h_{i} \left( n_{1}\right) $ and at $n_{1}^{\prime }$ player $i$ takes the action $a_{i}$.

\item[I7] Information sets consistent with own payoff information: For any $i \in I$, if $h_i(z) \subseteq T$ then $h_i(z) \subseteq Z^{T}$. Moreover, if $z' \in h_i(z)$ then $u_i(z') = u_i(z)$.

\end{itemize}

(I7) is new. It makes information sets of terminal nodes akin to feedback functions in the literature on self-confirming equilibrium. At any terminal node, a player considers only terminal nodes. That is, she knows that the game ended. Moreover, any two terminal nodes that a player cannot distinguish must yield her the same payoff because otherwise she could use her payoffs to distinguish among these terminal nodes. This implies that at the end of the game each player knows her own payoff. Note that this assumption does not rule out imperfect observability of \emph{opponents'} payoffs. It also does not rule out that the player may not perfectly observe the terminal node.

We denote by $H_{i}$ the set of $i$'s information sets in all trees. For an information set $h_{i} \in H_{i}$, we denote by $T_{h_{i}}$ the tree containing $h_{i}$. For two information sets $h_{i}, h_{i}^{\prime}$ in a given tree $T,$ we say that $h_{i}$ precedes $h_{i}^{\prime }$ (or that $h_{i}^{\prime }$ succeeds $h_{i}$) if for every $n^{\prime} \in h_{i}^{\prime }$ there is a path $n,...,n^{\prime }$ in $T$ such that $n \in h_{i}$. We denote it by $h_{i}\rightsquigarrow h_{i}^{\prime }$.

If $n \in h_i$ we write also $A_{h_i}$ for $A^i_{n}$.

To model unawareness proper, we impose as in Heifetz, Meier, and Schipper (2013) additional properties.

\begin{itemize}
\item[U4] Subtrees preserve ignorance: If $T \preceq T' \preceq T''$, $n \in T''$, $h_{i}(n) \subseteq T$ and $T'$ contains the copy $n_{T'}$ of $n$, then $h_{i}(n_{T'}) = h_{i}( n)$.

\item[U5] Subtrees preserve knowledge: If $T \preceq T' \preceq T''$, $n \in T''$, $h_{i}(n) \subseteq T'$ and $T$ contains the copy $n_{T}$ of $n$, then $h_{i}(n_{T})$ consists of the copies that exist in $T$ of the nodes of $h_{i}(n)$.
\end{itemize}

For trees $T, T^{\prime } \in \mathbf{T}$ we denote $T \rightarrowtail T^{\prime }$ whenever for some node $n\in T$ and some player $i\in P(n)$ it is the case that $h_{i}(n) \subseteq T'$. Denote by $\hookrightarrow $\ the transitive closure of $\rightarrowtail$. That is, $T \hookrightarrow T^{\prime \prime}$ if and only if there is a sequence of trees $T, T^{\prime }, \dots, T^{\prime \prime } \in \mathbf{T}$ satisfying $T \rightarrowtail T^{\prime} \rightarrowtail \dots \rightarrowtail T^{\prime\prime}$.

An \emph{extensive-form game with unawareness} $\Gamma $ consists of a join-semilattice $\mathbf{T}$ of subtrees of a tree $\bar{T}$ satisfying properties 1--3 above, along with information sets $h_i(n)$ for every $n \in T$ with $T \in \mathbf{T}$ and $i \in P(n)$, and payoffs satisfying properties U0, U1, U4, U5, and I2-I7 above.

For every tree $T \in \mathbf{T}$, the $T$\emph{-partial game} is the join-semisublattice of trees including $T$ and all trees $T^{\prime}$ in $\Gamma$ satisfying $T \hookrightarrow T^{\prime}$, with information sets as defined in $\Gamma$. A $T$-partial game is a extensive-form game with unawareness, i.e., it satisfies all properties 1--3, U0, U1, U4, U5, and I2-I7 above.

We denote by $H_{i}^{T}$ the set of $i$'s information sets in the $T$-partial game, $T \in \mathbf{T}$. This set contains not only $i$'s information sets in the tree $T$ but also in all trees $T' \in \mathbf{T}$ with $T \hookrightarrow T'$.

Further, we denote by $H_i^{\mathbf{D}}$ ($H_i^{T, \mathbf{D}}$, resp.) the set of $i$'s information sets of decision nodes (in the $T$-partial game, resp.) and by $H_i^{\mathbf{Z}}$ ($H_i^{T, \mathbf{Z}}$, resp.) the set of $i$'s information sets of terminal nodes (in the $T$-partial game, resp.).

\subsection{Strategies}

For any collection of sets $(X_i)_{i \in I^0}$ we denote by
$$X = \prod_{i \in I^0} X_{i}, \quad X_{-i} = \prod_{j \in I^0 \setminus \{i\}} X_{j}, \quad  X_{-i0} = \prod_{j \in I \setminus \{i\}} X_{j}$$ with typical elements $x$, $x_{-i}$, and $x_{-i0}$, respectively. For any collection of sets $(X_i)_{i \in I^0}$ and any tree $T \in \mathbf{T}$, we denote by $X_i^T$ the set of objects in $X_i$ restricted to the tree $T$ and analogously for $X^T$, $X^T_{-i}$, and $X^T_{-i0}$, where ``restricted to the tree $T$'' will become clear from the definitions below.

A \emph{pure strategy} for player $i$
\begin{equation*}
s_{i} \in S_{i} := \prod_{h_{i}\in H_{i}^{\mathbf{D}}} A_{h_{i}}
\end{equation*}
specifies an action of player $i$ at each of her information sets $h_{i}\in H_{i}^{\mathbf{D}}$ of decision nodes. We let
\begin{equation*}
s_0 \in S_0 := \prod_{n \in \mathbf{D}_0} A^0_{n}
\end{equation*} denote the ``strategy'' of nature, with $D_0$ denoting the ``decision'' nodes of nature.

With the strategy $s_{i}$, at node $n \in D_{i}^{T_n}$ define player $i$'s action at $n$ to be $s_i(h_{i}(n))$, for $i \in I$. Thus, by U1 and I4 the strategy $s_{i}$ specifies what player $i \in I$ does at each of her active nodes $n \in D_{i}^{T_n}$, both in the case that $n \in h_i(n)$ \emph{and} in the case that $h_i(n) $ is a subset of nodes of a tree which is distinct from the tree $T_{n}$ to which $n$ belongs. In the first case, when $n \in h_i(n)$, we can interpret $s_i(h_i(n))$ as the action chosen by player $i$ in node $i$. In the second case, when $n \notin h_i(n)$, $s_i(h_i(n))$ cannot be interpreted as the action chosen ``consciously'' by player $i$ in $n$ since he is not even aware of $T_n$. Instead, his state of mind at $n$ is given by his information set $h_i(n)$ in a tree lower than $T_n$ (denoted by $T_{h_i}$). Thus, $s_i(h_i(n))$ is the physical move of player $i$ in $n$ in tree $T_n$ induced by his ``consciously'' chosen action at his information set $h_i(n)$ in tree $T_{h_i(n)}$ (with $T_n \succ T_{h_i(n)}$). As an example, consider player 1 in the game of Figure~\ref{example1a}. At his first decision node in the upper tree $\bar{T}$, the root of the tree, player 1's information set consists of the corresponding node in the lower tree $T$. the strategy of player 1 may assign $r_1$ to his information set in the lower tree $T$. But it also induces action $r_1$ at the root of the upper tree $\bar{T}$.

In an extensive-form game with unawareness $\Gamma$ the tree $\bar{T} \in \mathbf{T}$ represents the physical paths in the game; every tree in $\mathbf{T}$ that contains an information set represents the subjective view of the feasible paths in the mind of a player, or the view of the feasible paths that a player believes that another player may have in mind, etc. Moreover, as the actual play in $\bar{T}$ unfolds, a player may become aware of paths of which she was unaware earlier, and the way she views the game may alter as well. Thus, in an extensive-form game with unawareness, a strategy cannot be conceived as an ex ante plan of action. Formally, a strategy of player $i$ is a list of answers to the questions ``what would player $i \in I$ do if $h_{i}$ were the set of nodes she considered as possible?'', for $h_{i}\in H_{i}$ (and analogous for nature). A strategy of a player becomes meaningful as an object of beliefs of other players. How ``much'' of a player's strategy other players can conceive off depend on their awareness given by the tree in which their information set is located. This leads to the notion of $T$-partial strategy. For a strategy $s_{i} \in S_{i}$ and a tree $T \in \mathbf{T}$, we denote by $s_{i}^{T}$ the strategy in the $T$-partial game induced by $s_{i}$ (i.e., $s_{i}^{T}\left( h_{i}\right) = s_{i}\left(h_{i}\right)$ for every information set $h_{i} \in H_i^T$ of player $i$ in the $T$-partial game).

A \emph{mixed strategy} of player $i$, $\sigma_{i} \in \Delta(S_{i})$, specifies a probability distribution over player $i$'s set of pure strategies. With this notation, we let $\sigma_0$ the probability distribution over ``strategies'' of nature. We don't consider mixed strategies as an object of choice of players; this notion will just be used in proofs in a technical way.

A \emph{behavioral strategy} for player $i \in I$,
\begin{equation*} \pi_i \in \Pi_i = \prod_{h_i \in H_i^{\mathbf{D}}} \Delta(A_{h_i})
\end{equation*}
is a collection of independent probability distributions, one for each of player $i$'s information sets $h_{i}\in H_{i}^{\mathbf{D}}$ of decision nodes, where $\pi_i(h_i)$ specifies a mixed action in $\Delta(A_{h_i})$. With the behavioral strategy $\pi_i$, at node $n \in D_i^{T_n}$ define player $i$'s mixed action at $n$ to be $\pi_i(h_i(n))$. Thus, the behavioral strategy $\pi_i$ specifies the mixed action of player $i \in I$ at each of her active decision nodes $n \in D_i^{T_n}$, both in the case that $n \in h_i(n)$ and in the case that $h_i(n)$ is a subset of nodes of a tree which is distinct from the tree $T_n$ to which $n$ belongs. In this latter case, we have automatically that $\pi_i$ does not assign probabilities to actions in $A_n \setminus A_{h_i(n)}$. (I.e., at the decision node $n$ of the richer tree $T_n$ player $i$ may have more actions than she is aware of at $h_i(n)$. In such a case, she is unable to use actions that she is unaware of.)

In extensive-form games with unawareness there are two distinct notions of a strategy profile being consistent with a node that we call ``reaching a node'' and ``visiting a node'', respectively. The difference between these two notions is relevant when we consider information sets that players believe are consistent with a strategy and information sets that are actually consistent with a strategy. Former is relevant for extensive-form rationalizability while latter is relevant for self-confirming equilibrium.

We say that a strategy profile $s=\left( s_{j}\right) _{j\in I}\in S$ \emph{reaches a node} $n\in T$ if the players' actions and nature's moves $\left(s_{j}^{T}\left( h_{j}\left( n^{\prime }\right) \right)\right)_{j \in P(n')}$ in nodes $n^{\prime }\in T$ lead to $n$. Notice that by property (I4) (``no imaginary actions''), $s_{j}^{T}\left( h _{j}\left( n^{\prime }\right) \right) _{j\in I}$ is indeed well defined: even if $h _{j}\left( n^{\prime }\right) \notin T$ for some $n^{\prime }\in T$, $\left(s_{j}^{T}\left( h_{j}\left( n'\right) \right)\right)_{j \in P(n')}$ is a profile of actions which is actually available in $T$ to the active players $j \in P(n')$ and possibly nature at $n'$. We say that a strategy profile $s\in S$ \emph{reaches} the information set $h_{i}\in H_{i}$ if $s$ reaches some node $n\in h_{i}$. We say that the strategy $s_{i}\in S_{i}$ \emph{reaches} the information set $h_{i}$ if there is a strategy profile $s_{-i}\in S_{-i}$ of the other players (and possibly nature) such that the strategy profile $\left( s_{i},s_{-i}\right)$ reaches $h_{i}$. Otherwise, we say that the information set $h_{i}$ is excluded by strategy $s_{i}$. Analogously, we say that the strategy profile $s_{-i} \in S_{-i}$ \emph{reaches} the information set $h_{i}$ if there exists a strategy $s_{i}\in S_{i}$ such that the strategy profile $\left( s_{i},s_{-i}\right)$ reaches $h_{i}$. For each player $i \in I$, denote by $H_i(s)$ the set of information sets of $i$ that are reached by the strategy profile $s$. This set typically contains information sets in more than one tree.

For the second notion of a strategy profile being consistent with a node, recall that a strategy $s_{i}$ specifies not only what player $i \in I$ does at nodes $n' \in h_i(n)$ but also in node $n$ where $n$ might be located in a tree more expressive than $T_{h_i(n)}$.

We say that a strategy profile $s=\left( s_{j}\right)_{j\in I}\in S$ \emph{visits a node} $n$ in the upmost tree $\bar{T}$ if the players' actions and nature's moves $\left(s_{j}\left( h_{j}(n') \right)\right)_{j \in P(n')}$ in nodes $n' \in \bar{T}$ lead to $n \in \bar{T}$. We extend the notion of a strategy profile visiting a node to any node in any tree by saying that a strategy profile $s=\left( s_{j}\right) _{j\in I}\in S$ \emph{visits a node} $n \in T$ if $s$ visits $n' \in \bar{T}$ with with $n'_T = n$. We say that a strategy profile $s\in S$ \emph{visits} the information set $h_{i}\in H_{i}$ if $s$ visits some node $n\in h_{i}$. We say that the strategy $s_{i}\in S_{i}$ \emph{visits} the information set $h_{i}$ if there is a strategy profile $s_{-i}\in S_{-i}$ of the other players (and possibly nature) such that the strategy profile $\left( s_{i},s_{-i}\right)$ visits $h_{i}$. Analogously, we say that the strategy profile $s_{-i} \in S_{-i}$ \emph{visits} the information set $h_{i}$ if there exists a strategy $s_{i}\in S_{i}$ such that the strategy profile $\left( s_{i},s_{-i}\right)$ visits $h_{i}$.

Define the \emph{path} $p(s, T)$ \emph{induced by strategy profile} $s$ in tree $T$ by the sequence of nodes in $T$ visited by $s$. Further, for any strategy profile $s$ and tree $T \in \mathbf{T}$, define $\tilde{H}_i(p(s, T)) := \{h_i \in H_i : h_i = h_i(n) \mbox{ for } n \mbox{ on the path } p(s, T) \mbox{ with } i \in P(n) \}$. Note that information sets in $\tilde{H}_i(p(s, T))$ may lie in different trees weakly less expressive than $T$.
\begin{figure}[h!]
\caption{Information Sets Visited versus Reached \label{infopath}}
\begin{center}
\includegraphics[scale=0.4]{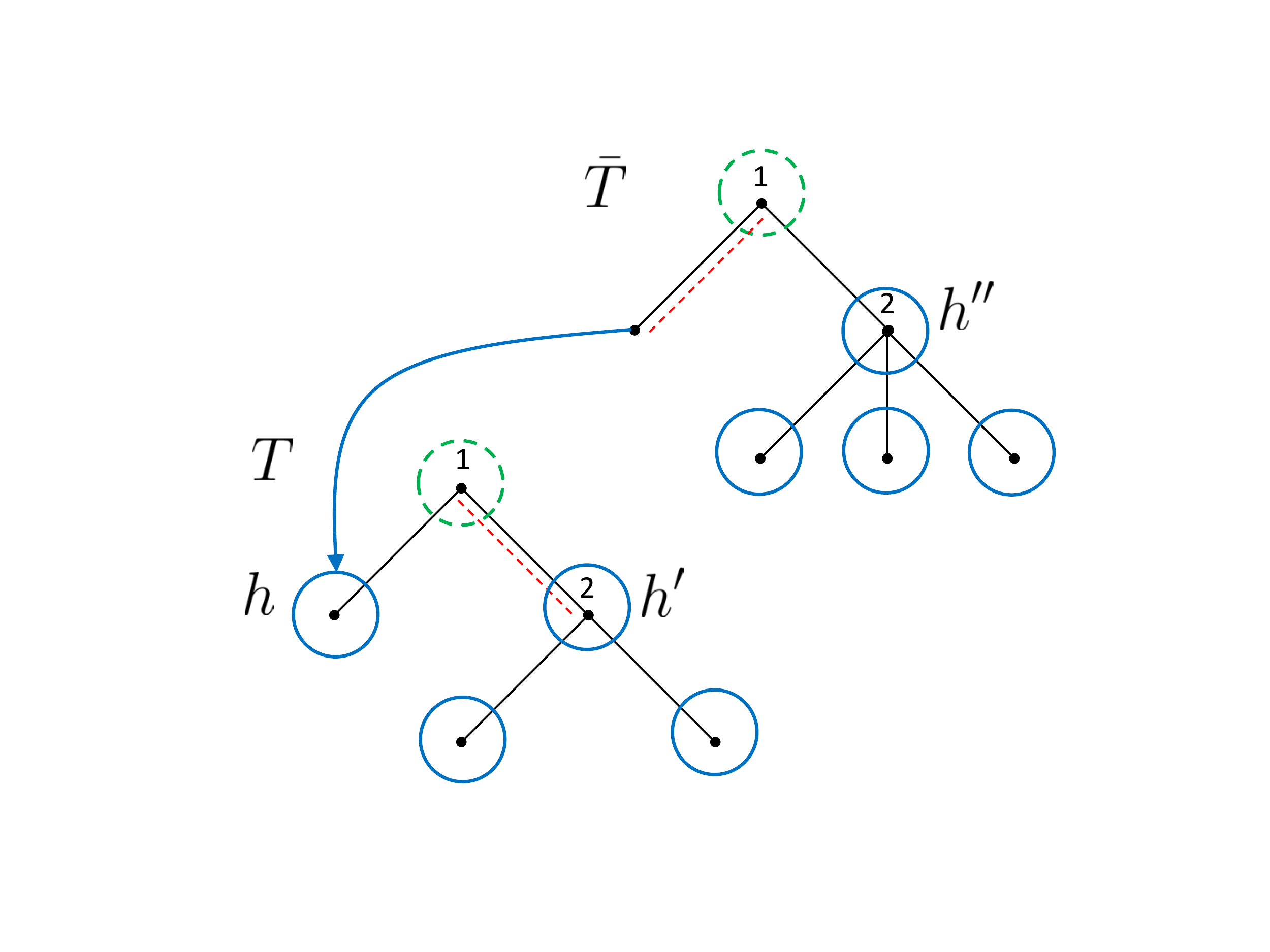}
\end{center}
\end{figure}

To clarify the subtle but important difference between the notions of visit and reach as well as the definitions of $\tilde{H}_i(p(s, T))$ and $H_i(s)$ consider the example in Figure~\ref{infopath}. There are two trees, $\bar{T} \succ T$. There are two players, 1 and 2. Player 1 moves first. As long as he moves right, player 2 moves second and the game ends. Otherwise the game ends. When player 1 moves left, player 2 remains unaware of her action middle. This is shown in Figure~\ref{infopath} by the blue arrows and ovals (i.e., information set $h$ in $T$) upon player 1 moving left. Otherwise, if player 1 moves right, player 2 becomes aware of middle (information set $h''$). (Player 1's initial information sets are indicated by green intermitted ovals.) Consider the strategy of player 1 indicated by the red intermitted edges. This strategy reaches $h'$ but visits $h$ in $T$. Information set $h''$ in tree $\bar{T}$ is neither reached nor visited by the strategy. Let $s$ denote any strategy profile in which player 1 follows the strategy indicated by the red intermitted edges. We have $h \in \tilde{H}_2(p(s, T))$, $h \in \tilde{H}_2(p(s, \bar{T}))$, $h, h'' \notin H_2(s)$, $h' \in H_2(s)$, $h', h'' \notin \tilde{H}_2(p(s, T))$, $h', h'' \notin \tilde{H}_2(p(s, \bar{T}))$.

It should be clear that in a standard extensive-form game (without unawareness), a strategy profile reaches a node $n$ if and only if it visits $n$.

We extend the definitions of information set reached and visited to behavioral strategies in the obvious way by considering nodes/information sets reached/visited with strict positive probability. Similarly, we let $p(\pi, T)$ denote now the set of paths that have strict positive probability under the behavioral strategy profile $\pi$ in $T$. $\tilde{H}_i(p(\pi, T))$ is now the set of information sets along paths in $p(\pi, T)$.

For any node $n$, any player $i \in I$, and any opponents' profile of strategies $s_{-i}$ (including nature if any), let $\rho(n \mid \pi_i, s_{-i})$ and $\rho(n \mid \sigma_i, s_{-i})$ denote the probability that $(\pi_i, s_{-i})$ and $(\sigma_i, s_{-i})$ reach node $n$, respectively. For any player $i \in I^0$, a mixed strategy $\sigma_i$ and a behavioral strategy $\pi_i$ are \emph{equivalent} if for every profile of opponents' strategies $s_{-i} \in S_{-i}$ and every node $n \in \mathbf{N}$ of the extensive-form game with unawareness $\rho(n \mid \sigma_i, s_{-i}) = \rho(n \mid \pi_i, s_{-i})$. Kuhn's Theorem can be extended to extensive-form games with unawareness (with perfect recall) so that for every player and for every mixed strategy of that player there exists an equivalent behavioral strategy.

\subsection{Belief Systems}

A \emph{belief system} of player $i$,
\begin{equation*}
\beta_{i}=\left( \beta_{i}\left( h_{i}\right) \right) _{h_{i}\in H_{i}} \in \prod_{h_{i}\in H_{i}}\Delta \left( S_{-i}^{T_{h_{i}}}\right)
\end{equation*}
is a profile of beliefs -- a belief $\beta_{i}\left( h_{i}\right) \in \Delta \left( S_{-i}^{T_{h_{i}}}\right)$ about the other players' strategies (and possibly nature) in the $T_{h_{i}}$-partial game, for each information set $h_{i}\in H_{i}$, with the following properties:

\begin{itemize}
\item $\beta_{i}\left( h_{i}\right)$ reaches $h_{i}$, i.e., $\beta_{i}\left( h_{i}\right) $ assigns probability 1 to the set of strategy profiles of the other players (including possibly nature) that reach $h_{i}$.

\item If $h_{i}$ precedes $h_{i}^{\prime }$ (i.e., $h_{i}\rightsquigarrow h_{i}^{\prime }$) then $\beta_{i}\left( h_{i}^{\prime} \right)$ is derived from $\beta_{i}\left( h_{i}\right)$ by Bayes rule whenever possible.
\end{itemize}

Note that different from Heifetz, Meier, and Schipper (2013) a belief system specifies also beliefs about strategies of opponents and nature at information sets of terminal nodes. This is an essentially feature that we will require for defining self-confirming equilibrium. Denote by $B_{i}$ the set of player $i$'s belief systems.

For a belief system $\beta_{i}\in B_{i}$, a strategy $s_{i}\in S_{i}$ and an information set $h_{i}\in H_{i},$ define player $i$'s expected payoff \emph{at} $h_{i}$ to be the expected payoff for player $i$ in $T_{h_{i}}$ given $\beta_{i}\left( h_{i}\right) $, the actions prescribed by $s_{i}$ at $h_{i}$ and its successors, assuming that $h_{i}$ has been reached.

We say that with the belief system $\beta_{i}$ and the strategy $s_{i}$ player $i$ is \emph{rational} at the information set $h_{i}\in H_{i}^{\mathbf{D}}$ if either $s_{i}$ does not reach $h_{i}$ or there exists no strategy $s_{i}^{\prime }$ which is distinct from $s_{i}$ only at $h_{i}$ and/or at some of $h_{i}$'s successors in $T_{h_{i}}$ and yields player $i$ a higher expected payoff in the $T_{h_{i}}$-partial game given the belief $\beta_{i}\left( h_{i}\right)$ on the other players' strategies $S_{-i}^{T_{h_{i}}}$.

Player $i$'s \emph{belief system on behavioral strategies} of opponents
\begin{eqnarray*}
\mu_i = \left(\mu_i(h_i)\right)_{h_i \in H_i} \in \prod_{h_i \in H_i} \Delta(\Pi_{-i}^{T_{h_i}})
\end{eqnarray*} is a profile of beliefs -- a belief $\mu_i(h_i) \in \Delta(\Pi_{-i}^{T_{h_i}})$ about the behavioral strategies of other players (incl. possibly nature) in the $T_{h_i}$-partial game, for each information set $h_i \in H_i$, with the following properties
\begin{itemize}
\item $\mu_{i}\left( h_{i}\right)$ reaches $h_{i}$, i.e., $\mu_{i}\left( h_{i}\right) $ assigns probability 1 to the set of behavioral strategy profiles of the other players (incl. possibly nature) that reach $h_{i}$.

\item If $h_{i}$ precedes $h_{i}^{\prime }$ (i.e., $h_{i}\rightsquigarrow h_{i}^{\prime }$) then $\mu_{i}\left( h_{i}^{\prime} \right)$ is derived from $\mu_{i}\left( h_{i}\right)$ by Bayes rule whenever possible.
\end{itemize} We denote by $M_i$ the set of player $i$'s belief systems over behavioral strategies of opponents.

For a belief system $\mu_{i}\in M_{i}$, a behavioral strategy $\pi_{i} \in \Pi_{i}$ and an information set $h_{i}\in H_{i}$, define player $i$'s expected payoff \emph{at} $h_{i}$ to be the expected payoff for player $i$ in $T_{h_{i}}$ given $\mu_{i}\left( h_{i}\right)$, the mixed actions prescribed by $\pi_{i}$ at $h_{i}$ and its successors, assuming that $h_{i}$ has been reached.

We say that with the belief system $\mu_{i}$ and the behavioral strategy $\pi_{i}$ player $i$ is \emph{rational} at the information set $h_{i}\in H_{i}^{\mathbf{D}}$ if either $\pi_{i}$ does not reach $h_{i}$ or there exists no behavioral strategy $\pi_{i}'$ which is distinct from $\pi_{i}$ only at $h_{i}$ and/or at some of $h_{i}$'s successors in $T_{h_{i}}$ and yields player $i$ a higher expected payoff in the $T_{h_{i}}$-partial game given the belief $\mu_{i}\left( h_{i}\right)$ on the other players' behavioral strategies $\Pi_{-i}^{T_{h_{i}}}$.

\subsection{Self-Confirming Equilibrium\label{sce_section}}

The discussion of the example made clear that the challenge for a notion of equilibrium is to deal with changes of awareness along the equilibrium paths. In a ``steady state of conceptions'', awareness should not change. We incorporate this requirements into our definition of self-confirming equilibrium.

\begin{defin}\label{sceb} A behavioral strategy profile $\pi \in \Pi$ is a self-confirming equilibrium if for every player $i \in I$:
\begin{itemize}
\item[(0)] Awareness is self-confirming along the path: There is a tree $T \in \mathbf{T}$ such that for all of player $i$'s visited information sets $h_i \in \tilde{H}_i(p(\pi, \bar{T}))$ we have $h_i \subseteq T$.
\end{itemize}
\noindent There exists a belief system\footnote{We do not require that player $i$ believes that opponents mix independently as this is hard to motivate. In the literature on self-confirming equilibrium independence is assumed in Fudenberg and Levine (1993) but not in Rubinstein and Wolinsky (1994).} $\mu_i \in M_i$ such that
\begin{itemize}
\item[(i)] Players are rational along the path: With belief system $\mu_i$, behavioral strategy $\pi_i$ is rational at all visited information sets in $\tilde{H}_i(p(\pi, \bar{T}))$.
\item[(ii)] Beliefs are self-confirming along the path: For the information set of terminal nodes $h_i \in H_i^{\mathbf{Z}} \cap \tilde{H}_i(p(\pi, \bar{T}))$ visited by the behavioral strategy profile $\pi$, the belief system $\mu_i$ is such that $\mu_i(h_i)$ assigns probability 1 to $\{\pi'_{-i} \in \Pi_{-i} : \pi'_j(h_j) = \pi_j(h_j)$ for $j \in I^0 \setminus \{i\}$ and $h_j \in \tilde{H}_j(p(\pi, T_{h_i}))\}$. Moreover, for any preceding (hence non-terminal) information set $h_i' \rightsquigarrow h_i$, $\mu_i(h_i') = \mu_i(h_i)$.
\end{itemize}
\end{defin}

Condition (0) requires that awareness is constant along the equilibrium path. Players do not discover anything novel in equilibrium play. This is justified by the idea of equilibrium as a stationary rest-point or stable convention of play. Implicitly, it is assumed that discoveries if any are made before equilibrium is reached.

Condition (i) is a basic rationality requirement of equilibrium. Note that rationality is required only along information sets that occur along the path of play induced by the equilibrium strategy profile. The equilibrium notion is silent on off-equilibrium information sets (in particular on information sets that could be visited with $s_i$ but are not visited with $s_{-i}$). Condition (i) does also not require that players believe others are rational along the path, believe that others believe that etc. It is just a ``minimal'' rationality requirement in an extensive-form game.

Condition (ii) consists of two properties. First, at the end of the game the player is certain of observationally equivalent behavioral strategies of opponents and nature that allow her to reach the particular end of the game. That is, terminal beliefs are consistent with what has been observed during play (and hence at the end of the play). Second, beliefs do not change during the play. That is, beliefs at any information set reached during the play are consistent with what is observed at any point during the play and in particular with what is observed at the end of the game. Again, the idea is that everything that could have been learned on this path has been learned already in the past. This is justified by the idea of equilibrium as a stationary rest-point or stable convention of play as a result of prior learning. Note that this notion of equilibrium is silent on beliefs off equilibrium path.

While it is well-known in the literature (see for instance Fudenberg and Levine, 1993, Fudenberg and Kreps, 1995, Battigalli and Guaitoli, 1997) that self-confirming equilibrium is a coarsening of Nash equilibrium, the point here is to contrast it with the case of unawareness. Despite the fact that self-confirming equilibria are a coarsening of Nash equilibria, they may not exist in finite extensive-form games with unawareness due to failure of condition (0). The game in Figure~\ref{example1a} constitutes a simple counterexample.\\

\noindent \textbf{Example (continued): Failure of self-confirming equilibrium in finite extensive-form games with unawareness.} Condition (i), rationality along the path, requires that player 1 chooses $\ell_1$ and player 2 chooses $m_2$ in $\bar{T}$ and $r_2$ in $T$ in the game of Figure~\ref{example1a}. It is also easy to see that for each player there exists a belief system satisfying condition (ii) of the definition of self-confirming equilibrium. Yet, the play emerging from rational strategies reaches player 1's information set containing a terminal node in $\bar{T}$ after being initially only aware of $T$, which violates condition (0). That is, awareness is not self-confirming along the path. Hence, there is no self-confirming equilibrium.\\

It should be clear that by introducing self-confirming equilibrium we strive for the weakest possible notion of equilibrium embodying the idea of steady-state in order to observe that even such weak notion may still not exist in finite games with unawareness due to changes of awareness. While this weak notion of equilibrium rules out changes of awareness, it does not necessarily imply mutual knowledge (or even common knowledge) of no changes of awareness, a condition that may be reasonably imposed in a satisfactory equilibrium concept for games with unawareness.

\section{Discovery Processes\label{discovery_section}}

Let $\bm{\Gamma}$ be the set of all extensive-form games with unawareness for which the initial building block is the finite extensive-form game with perfect information $\langle I, \bar{T}, P, (u_i)_{i \in I}\rangle$. By definition, $\bm{\Gamma}$ is finite.

For any extensive-form game with unawareness $\Gamma \in \bm{\Gamma}$, denote by $S_{\Gamma}$ the set of pure strategy profiles in $\Gamma$.

\begin{defin}\label{discovered_version} Given an extensive-form game with unawareness $\Gamma = \langle I, \mathbf{T}, P, (H_i)_{i \in I}, (u_i)_{i \in I} \rangle \in \bm{\Gamma}$ and a strategy profile in this game $s_{\Gamma}$, the discovered version $\Gamma' = \langle I', \mathbf{T}', P', (H_i')_{i \in I'}, (u'_i)_{i \in I'} \rangle$ is defined as follows:
\begin{itemize}
\item[(i)] $I' = I$, $\mathbf{T}' = \mathbf{T}$, $P' = P$, and $u'_i = u_i$ for all $i \in I'$.

\item[(ii)] For $i \in I'$, the information sets in $H_i'$ of $\Gamma'$ are defined as follows: Let $$T^i_{s_{\Gamma}} := \sup\left\{T \in \mathbf{T} : h_i(n) \subseteq T, h_i(n) \in \tilde{H}_i(p(s_{\Gamma}, \bar{T}))\right\}.$$

    For any $n \in \bar{T}$ with $h_i(n) \in H_i$, $h_i(n) \subseteq T'$, $T', T'' \in \mathbf{T}$,
    \begin{itemize}
    \item[a.] if $T' \preceq T^i_{s_{\Gamma}} \preceq T'' \preceq \bar{T}$, the information set $h_i'(n_{T''}) \in H_i'$ is defined by\footnote{As defined previously, we take $n_{T''}$ the copy of node $n \in \bar{T}$ in the tree $T''$. When $T'' \equiv \bar{T}$, then $n_{T''} = n$.}
        $$h_i'(n_{T''}) := \left\{n' \in T^i_{s_{\Gamma}}: h_i(n') = h_i(n)\right\}.$$

    \item[b.] if $T' \preceq T'' \preceq T^i_{s_{\Gamma}}$, the information set $h_i'(n_{T''}) \in H_i'$ is defined by $$h_i'(n_{T''}) := \left\{n' \in T'': h_i(n') = h_i(n)\right\}.$$

    \item[c.] Otherwise, if $T' \not\preceq T^i_{s_{\Gamma}}$ and $T' \preceq T'' \preceq \bar{T}$, the information set $h_i'(n_{T''}) \in H_i'$ is defined by $$h_i'(n_{T''}) := h_i(n_{T''}).$$

    \end{itemize}

\end{itemize}
\end{defin}

When an extensive-form game with unawareness $\Gamma$ is played according to a strategy profile $s_{\Gamma}$, then some players may discover something that they were previously unaware of. The discovered version $\Gamma'$ of an ``original'' extensive-form game with unawareness $\Gamma$ represents the views of the players after the extensive-form game with unawareness has been played according to a strategy profile $s_{\Gamma}$. The discovered version has the same set of players, the same join-semilattice of trees, the same player correspondence, and the same payoff functions as the original game. What may differ are the information sets. In particular, in a discovered version some players may from the beginning be aware of more actions than in the ``original'' game $\Gamma$ but only if in the ``original'' game it was possible to discover these action with the strategy profile $s_{\Gamma}$. The information sets in the discovered version reflect what players have become aware of when playing $\Gamma$ according to $s_{\Gamma}$.

To understand part $(ii)$ of Definition~\ref{discovered_version}, note first that $T^i_{s_{\Gamma}}$ is the tree that represents the player $i$'s awareness of physical moves in the game $\Gamma$ after it has been played according to strategy profile $s_{\Gamma}$. It is determined by the information sets of player $i$ that occur along the play-path in the upmost tree according to $s_{\Gamma}$. Now consider all information sets of player $i$ that arise at nodes in the upmost tree $\bar{T}$ in the ``original'' game $\Gamma$. These information sets may be on lower trees than $T^i_{s_{\Gamma}}$. Since player $i$ is now aware of $T^i_{s_{\Gamma}}$, all those information sets that in $\Gamma$ were on a tree lower than $T^i_{s_{\Gamma}}$ are now lifted to tree $T^i_{s_{\Gamma}}$, the tree that in player $i$'s mind represents the physical moves of the strategic situation after $\Gamma$ has been played according to $s_{\Gamma}$. Yet, this holds not only for reached nodes in the upmost tree $\bar{T}$ but also for copies of those nodes in trees $T'' \in \mathbf{T}$, $T^i_{s_{\Gamma}} \preceq T'' \preceq \bar{T}$. This is because by Property U4 of extensive-form games with unawareness the information sets at copies of those nodes in trees $T''$ are also on trees lower than $T^i_{s_{\Gamma}}$ in $\Gamma$. When information sets are lifted to higher trees, they contain all nodes in such a tree that previously gave rise to the information set at a lower tree in $\Gamma$. This explains part $a.$ of $(ii)$ of Definition~\ref{discovered_version}.

Part $b.$ pertains to information sets in trees below $T^i_{s_{\Gamma}}$. These are trees that miss certain aspects that player $i$ is aware of in tree $T^i_{s_{\Gamma}}$. These trees are relevant to player $i$ nevertheless as she has to consider other player's views of the strategic situation, their views of her view etc. Since other players may be unaware of aspect that player $i$ is aware of, she should consider her own ``incarnations'' with less awareness. In the discovered version, the information sets on trees $T'' \preceq T^i_{s_{\Gamma}}$ model the same knowledge of events as in information sets on tree $T^i_{s_{\Gamma}}$ provided that she is still aware of those events in $T''$. This is crucial for the discovered version to satisfy property U5 of extensive-form games with unawareness.

Part $c.$ just says that in the discovered version information set of player $i$ in trees incomparable to $T^i_{s_{\Gamma}}$ remain identical to the original game $\Gamma$. These information sets represent awareness that necessarily has not been discovered when $\Gamma$ is played according to strategy profile $s_{\Gamma}$.
\begin{figure}[h!]
\caption{Original (left game form) and Discovered Version (right game form)\label{discovered}}
\begin{center}
\includegraphics[scale=0.5]{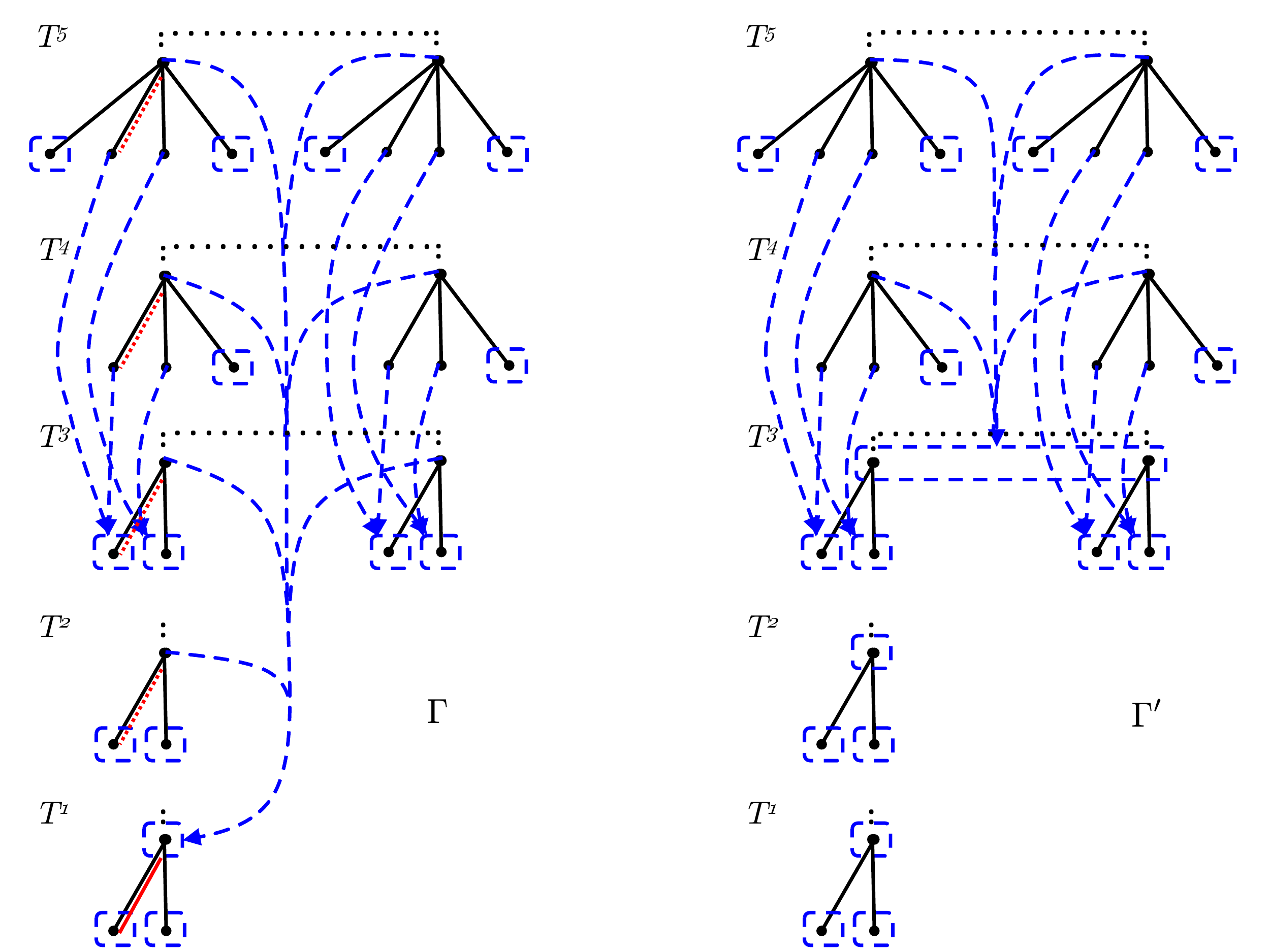}
\end{center}
\end{figure}

The notion of discovered version is illustrated in Figure~\ref{discovered}. This example is sufficiently rich to cover at least cases a. and b. distinguished in (ii) of Definition~\ref{discovered_version}. Consider the extensive-form game with unawareness to the left, $\Gamma$, as the ``original'' game. The extensive-form games with unawareness to the right is the discovered version $\Gamma'$ if players (and possibly nature) follow the strategy profiles indicated by the orange dashed lines. Clearly, $T^5$ in Figure~\ref{discovered} corresponds to $\bar{T}$ in Definition~\ref{discovered_version} (ii), $T^1$ to $T'$, and $T^3$ to $T^i_{s_{\Gamma}}$. For case a., let $T''$ in Definition~\ref{discovered_version} (ii) be $T^4$. For case b., let $T''$ correspond to $T^2$.

It is intuitive that when a player discovers something, her awareness is raised. Consequently, discovered versions of a game involve more awareness. The following definition makes this precise.

\begin{defin} Consider two extensive-form games with unawareness $\Gamma = \langle I, \mathbf{T}, P, (H_i)_{i \in I}, (u_i)_{i \in I} \rangle$ and $\Gamma' =$ \\ $\langle I', \mathbf{T}', P', (H_i')_{i \in I'}, (u'_i)_{i \in I'} \rangle$ with $I' = I$, $\mathbf{T}' = \mathbf{T}$, $P' = P$, and $u'_i = u_i$ for all $i \in I'$. $\Gamma'$ has (weakly) more awareness than $\Gamma$ if for every node $n$ and every active player $i \in P(n)$, $h_i(n) \subseteq T$ and $h'_i(n) \subseteq T'$ implies $T' \succeq T$.
\end{defin}

Discovered versions shall just reflect changes of awareness. The information about play, i.e., what players know about the history in the game, should not change. That is, in a discovered version players should have the same knowledge or ignorance about play modulo awareness as in the original game. The following definition makes this precise.

\begin{defin} Consider two extensive-form games with unawareness $\Gamma = \langle I, \mathbf{T}, P, (H_i)_{i \in I}, (u_i)_{i \in I} \rangle$ and $\Gamma' =$ \\ $\langle I', \mathbf{T}', P', (H_i')_{i \in I'}, (u'_i)_{i \in I'} \rangle$ with $I' = I$, $\mathbf{T}' = \mathbf{T}$, $P' = P$, and $u'_i = u_i$ for all $i \in I'$ such that $\Gamma'$ has (weakly) more awareness than $\Gamma$. $\Gamma'$ preserves information of $\Gamma$ if
\begin{itemize}
\item[(i)] for any $n$ and every active player $i \in P(n)$, $h_i(n)$ consists of copies of nodes in $h'_i(n)$.
\item[(ii)] for any tree $T \in \mathbf{T}$, any two nodes $n, n' \in T$ and every active player $i \in P(n) \cap P(n')$, if $h_i(n) = h_i(n')$ then $h'_i(n) = h'_i(n')$.
\end{itemize}
\end{defin}

Discovered versions are well-defined, have weakly more awareness than the original game and preserve information of the original game.

\begin{prop}\label{well-defined} For any extensive-form game with unawareness $\Gamma \in \bm{\Gamma}$ and any strategy profile $s_{\Gamma} \in S_{\Gamma}$ in this game, the discovered version $\Gamma'$ is an extensive-form game with unawareness. Moreover, $\Gamma'$ has more awareness than $\Gamma$. Finally, $\Gamma'$ preserves the information of $\Gamma$.
\end{prop}

Discovered versions do not depend on differences in strategies that are irrelevant to discoveries. They only depend on the realized path. This follows immediately from Definition~\ref{discovered_version} because $T^i_{s_\Gamma}$ depends only on $\tilde{H}_i(p(s_{\Gamma}, \bar{T}))$.

The players interaction may lead to discoveries, interaction in the discovered games may lead to further discoveries etc. To model the set of discovery processes based on an extensive-form game with unawareness, we essentially define a stochastic game in which each state represents an extensive-form game with unawareness.

\begin{defin} The discovery game based on $\bm{\Gamma}$ is the stochastic game $\langle \bm{\Gamma}, \tau \rangle$ defined as follows
\begin{itemize}

\item the set of states is a finite set of all extensive-form games with unawareness $\bm{\Gamma}$ (with identical initial building block $\langle I, \bar{T}, P, (u_i)_{i \in i} \rangle$.)

\item the transition probabilities are given by for $\Gamma, \Gamma' \in \bm{\Gamma}$, $s \in S_{\Gamma}$,
    $$\tau(\Gamma' \mid \Gamma, s) = \left\{\begin{array}{cl} 1 & \mbox{if } \Gamma' \mbox{ is the discovered version of } \\ &  \Gamma \mbox{ given } s \\
    0 & \mbox{otherwise} \end{array}\right.$$

\end{itemize}
\end{defin}

The discovery game is a stochastic game in which states are identified with extensive-form games with unawareness all based on the same building blocks. This means that each stage game is an extensive-form game with unawareness. The set of players is the set of players in the underlying extensive-form games with unawareness (including nature if any). Each player's set of actions is state-dependent and consist of the strategies at the extensive-form games with unawareness. The transition probabilities are degenerate in the sense that only transitions to discovered versions are allowed (given the strategy profiles). Payoffs of players are given by the underlying extensive-form games with unawareness. Since all those games have the same building blocks, payoffs are in fact the same in all states. What changes from state to state are information sets (and hence the set of strategies available at those stage games).

Clearly, the discovery game cannot be interpreted as a game that players are aware of. Rather, it is a convenient model for the modeler/analyst. Consequently, the supergame strategies of player in this discovery game are not objects actually chosen by players but just conveniently summarize the modeler's belief about player's play in all those games. A \emph{discovery game strategy} of player $i$ in the discovery game $\langle \bm{\Gamma}, \tau \rangle$ is a mapping $f_i: \bm{\Gamma} \longrightarrow \bigcup_{\Gamma \in \bm{\Gamma}} \Delta\left(S_{\Gamma, i}\right)$ that assigns to each game $\Gamma \in \bm{\Gamma}$ a probability distribution over strategies of that player in this game $\Gamma$ (i.e., $f_i(\Gamma) \in \Delta(S_{\Gamma, i})$). The notion makes clear that we only consider discovery strategies that are stationary Markov strategies. For each player $i \in I^0$ (including nature) denote by $F_i$ the set of all discovery strategies and by $F = \times_{i \in I^0} F_i$. Denote by $f = (f_i)_{i \in I^0}$ a profile of discovery strategies. We extend the definition of transition probabilities in order to be able to write $\tau(\cdot \mid \Gamma, f)$ for any $\Gamma \in \bm{\Gamma}$ and $f \in F$.

\begin{defin} A discovery process $\langle \bm{\Gamma}, \tau, (f_i)_{i \in I^0} \rangle$ consists of a discovery game $\langle \bm{\Gamma}, \tau \rangle$ and a discovery strategy $f_i: \bm{\Gamma} \longrightarrow \bigcup_{\Gamma \in \bm{\Gamma}} \Delta\left(S_{\Gamma, i}\right)$, one for each player $i \in I^0$ (including nature if any).
\end{defin}

In our formulation, every discovery process is a Markov process. An extensive-form game with unawareness $\Gamma \in \bm{\Gamma}$ is an absorbing state of the discovery process $\langle \bm{\Gamma}, \tau, f \rangle$ if $\tau(\Gamma \mid \Gamma, f) = 1$.

\begin{defin} An extensive-form game with unawareness $\Gamma \in \bm{\Gamma}$ is a self-confirming game of a discovery process $\langle \bm{\Gamma}, \tau, f \rangle$ if $\Gamma$ is an absorbing state of $\langle \bm{\Gamma}, \tau, f \rangle$.
\end{defin}

This terminology is justified by the fact in a self-confirming game play won't lead to further discoveries and changes of awareness and the information structure. All players' subjective representations of the game are in a steady-state. In this sense, the game is self-confirming. It is easy to find examples of self-confirming games in which players do no have common constant awareness.

\begin{prop}\label{absorbing} Every discovery process leads to a self-confirming game.
\end{prop}

It is easy to come up with examples in which a discovery process has more than one self-confirming game.

\section{Rationalizable Discoveries\label{rat_discovery_section}}

How to select among discovery processes? Which behavioral assumptions should be imposed on discovery processes? Clearly, it would be absurd to assume that players chose optimal discovery strategies since this would presume awareness of the discovery game and hence awareness of everything modelled in $\bm{\Gamma}$. In other words, there wouldn't be anything to discover.

We propose to restrict discovery processes to extensive-form rationalizable strategies (Pearce, 1984, Battigalli, 1997, Heifetz, Meier, and Schipper, 2013). A rational player in a novel game should be able to reason about the rationality of others, their (strong) beliefs about rationality etc. It is easy to find examples in which restricting discovery processes to extensive-form rationalizable strategies effectively selects among discovery processes and games that can be discovered in such processes.

\begin{defin}\label{EFR} Define, inductively, the following
sequence of belief systems and strategies of player $i \in I$
\begin{eqnarray*}
B_{i}^{1} & = & B_{i} \\
R_{i}^{1} & = & \left\{ s_{i}\in S_{i}:
\begin{array}{l}
\mbox{there exists a belief system } \beta_{i} \in B_{i}^{1} \\
\mbox{with which for every information set} \\
h_{i} \in H_{i} \mbox{ player } i \mbox{ is rational at } h_{i}
\end{array}
\right\} \\
& \vdots & \\
B_{i}^{k} & = & \left\{\beta_{i} \in B_{i}^{k-1}:
\begin{array}{l}
\mbox{for every information set } h_{i}, \\
\mbox{if there exists some profile } \\
\mbox{of the other players' strategies } \\
s_{-i} \in R_{-i}^{k-1} = \prod_{j \neq i} R_{j}^{k-1} \mbox{ such} \\
\mbox{that } s_{-i} \mbox{ reaches } h_{i}, \mbox{ then } \beta_{i}(h_{i}) \\
\mbox{ assigns probability } 1 \mbox{ to } R_{-i}^{k-1, T_{h_{i}}}
\end{array}
\right\} \\
R_{i}^{k} & = & \left\{ s_{i} \in S_{i}:
\begin{array}{l}
\mbox{there exists a belief system } \pi_{i}\in \Pi_{i}^{k} \\
\mbox{with which for every information set } \\
h_{i} \in H_{i} \mbox{ player } i \mbox{ is rational at } h_{i}
\end{array}
\right\}
\end{eqnarray*}
The set of player $i$'s extensive-form rationalizable strategies is
\begin{equation*}
R_{i}^{\infty } = \bigcap_{k=1}^{\infty} R_{i}^{k}.
\end{equation*}
\end{defin}

Denote by $R^{\infty}_{\Gamma, i}$ the set of extensive-form rationalizable strategies of player $i$ in the extensive-form game with unawareness $\Gamma$. A rationalizable discovery process is now defined as a discovery process where for each extensive-form game with unawareness each player is restricted to play extensive-form rationalizable strategies only.

\begin{defin} A discovery process $\langle \bm{\Gamma}, \tau, (f_i)_{i \in I^0} \rangle $ is a rationalizable discover process if for all players $i \in I$, $f_i: \bm{\Gamma} \longrightarrow \bigcup_{\Gamma \in \bm{\Gamma}} \Delta\left(R^{\infty}_{\Gamma, i}\right)$.
\end{defin}

Often, the analyst wants to analyze a particular game with unawareness. Thus, it will be helpful to designate it as the initial state of the discovery game. We denote by $\langle \bm{\Gamma}, \tau, \Gamma^0 \rangle$ the discovery game with \emph{initial game} $\Gamma^0$.

The next proposition shows that for every extensive-form game with unawareness $\Gamma^0$ there exists a rationalizable self-confirming version.

\begin{prop}\label{existence_rscv} For every extensive-form game with unawareness $\Gamma^0$ there exists a rationalizable discovery process $\langle \bm{\Gamma}, \tau, \Gamma^0, (f_i) \rangle$ that leads to a self-confirming game. We call such a self-confirming game a rationalizable self-confirming game.
\end{prop}

\section{Equilibrium\label{equilibrium}}

Previously we argued that often extensive-form games with unawareness do not possess equilibria that capture the result of a learning process because of the self-destroying nature of games with unawareness. Yet, since for every extensive-form game with unawareness there exists a discovery process that leads to a self-confirming version, the appropriate notion of equilibrium of a game with unawareness should naturally involve the equilibrium in the self-confirming version. Moreover, we also argued to restrict discovery processes to rationalizable discovery processes. This motivates to restrict equilibria to rationalizable strategies as well since it would be odd to assume that players play extensive-form rationalizable strategies all along the discovery process but once a rationalizable self-confirming version is reached, their equilibrium play might involve strategies that are not extensive-form rationalizable. That is, we propose to use extensive-form rationalizability not only in order to put endogenously restrictions on the games that can be discovered but also on the self-confirming equilibrium that may emerge in final states of discovery processes. While self-confirming equilibrium is a rather weak solution concept, the requirement of using only extensive-form rationalizable strategies strengthens it considerably as extensive-form rationalizability involves forward induction.

\begin{defin}\label{rsce} A behavioral strategy profile $\pi^* = (\pi_i^*)_{i \in I^0} \in \Pi$ is a rationalizable self-confirming equilibrium of the extensive-form game with unawareness $\Gamma$ if for every player $i \in I$ there is a mixed strategy $\sigma^*_i$ equivalent to $\pi^*_i$ that assigns zero probability to every strategy of player $i$ that is not extensive-form rationalizable.
\end{defin}

Rationalizable self-confirming equilibrium refines self-confirming equilibrium.

The following theorem asserts that for every finite extensive-form game with unawareness there exists a steady state of conceptions and behavior emerging from rationalizable play.

\begin{theo}\label{main_theorem} For every extensive-form game with unawareness there exists a rationalizable discovery process leading to a rationalizable self-confirming game which possesses a rationalizable self-confirming equilibrium.
\end{theo}


\bibliographystyle{eptcs}

\end{document}